\documentclass[runningheads]{llncs}
\usepackage{graphicx}
\begin{document}
\title{3D Convolutional Neural Networks for Ultrasound-Based Silent Speech Interfaces}
\titlerunning{3D CNNs for Silent Speech Interfaces}
%
\author{László Tóth \and
Amin Honarmandi Shandiz}
%
\authorrunning{L. Tóth and A. Shandiz.}
%
\institute{University of Szeged, Institute of Informatics\\
\email{\{tothl, shandiz\}@inf.u-szeged.hu}
}
\maketitle              

\begin{abstract}
Silent speech interfaces (SSI) aim to reconstruct the speech signal from
a recording of the articulatory movement, such as an ultrasound video of the tongue. Currently, deep neural networks are the most successful technology for this task. The efficient solution requires methods that do not simply process single images, but are able to extract the tongue movement information from a sequence of video frames. One option for this is to apply recurrent neural structures such as the long short-term memory network (LSTM) in combination with 2D convolutional neural networks (CNNs). Here, we experiment with another approach that extends the CNN to perform 3D convolution, where the extra dimension corresponds to time. In particular, we apply the spatial and temporal convolutions in a decomposed form, which proved very successful recently in video action recognition. We find experimentally that our 3D network outperforms the CNN+LSTM model, indicating that 3D CNNs may be a feasible alternative to CNN+LSTM networks in SSI systems.

\end{abstract}
\section{Introduction}

During the last couple of years, there has been an increasing interest in articulatory-to-acoustic conversion, which seeks to reproduce the speech signal from a recording of the articulatory organs, giving the technological background
for creating “Silent Speech Interfaces” (SSI) \cite{Denby2010,Schultz2017a}. These interfaces allow us to record the soundless articulatory movement, and then automatically generate speech from the movement information, while the subject is actually not producing any sound. Such an SSI system could be very useful for the speaking impaired who are able to move their articulators, but have lost their ability to produce any sound (e.g. due to a laryngectomy or some injury of the vocal chords). It could also be applied in human-computer interaction in situations where regular speech is not feasible (e.g. extremely noisy environments or military applications). Several solutions exist for the recording of the articulatory movements, the simplest approach being a lip video \cite{ephrat2017vid2speech,akbari2018lip2audspec}. But one may also apply 
electromagnetic articulography (EMA, \cite{Kim2017,Kim2017a}), ultrasound tongue imaging (UTI, \cite{Jaumard-Hakoun2016,Csapo2017c,Grosz2018,SottoVoce}) or permanent magnetic articulography (PMA, \cite{Gonzalez2017a}). Surface Electromiography (sEMG, \cite{Maier-Hein2005,Janke2012,Janke2017}) is also an option, while some authors use a combination of the above methods \cite{Denby2010}. Here we are going to work with ultrasound tongue videos.

To convert the movement recordings into speech, the conventional approach is to apply a two-step procedure of 'recognition-and-synthesis'~\cite{Schultz2017a}. In this case, the biosignal is first converted into text by a properly adjusted speech recognition system. The text is then converted into speech using text-to-speech synthesis~\cite{Denby2011,Hueber2010,Wang2014}. The drawbacks of this approach are the relatively large delay between the input and the output, and that the errors made by the speech recognizer will inevitably appear as errors in the TTS output. Also, all information related to speech prosody is lost, while certain prosodic components such as energy and pitch can be reasonably well estimated from the articulatory signal~\cite{Grosz2018}.

Current SSI systems prefer the `direct synthesis' principle, where speech is generated directly from the articulatory data, without any intermediate step. 
Moreover, as recently the Deep Neural Network (DNN) technology have become dominant in practically all areas of speech technology, such as speech recognition~\cite{Hinton2012}, speech synthesis~\cite{Ling2015} and language modeling~\cite{young_nlp}, most recent studies have attempted to solve the articulatory-to-acoustic conversion problem by using deep learning, regardless of the recording technique applied~\cite{Csapo2017c,Grosz2018,Janke2017,Jaumard-Hakoun2016,Gonzalez2017a,Juanpere2019,SottoVoce}. In this paper, we also apply deep neural networks to convert the ultrasound video of the tongue movement to speech. Although some early studies used simple fully connected neural networks~\cite{Jaumard-Hakoun2016,Csapo2017c}, as we are working with images, it seems more reasonable to apply convolutional neural networks (CNN), which are currently very popular and successful in image recognition~\cite{Imagenet}. Thus, many recent studies on SSI systems use CNNs~\cite{Janke2017,Juanpere2019,SottoVoce}.

Our input here is a video, that is, not just one still image, but a sequence of images. This sequence carries extra information about the time trajectory of the tongue movement, which might be exploited by processing several neighboring video frames at the same time. There are several options to create a network structure for processing a time sequences. For such data, usually recurrent neural networks such as the long short-term memory network (LSTM) are applied, typically stacking it on top of a 2D CNN that seeks to process the individual frames~\cite{Gonzalez2017a,Juanpere2019,Liu2018,Kim2017a}. Alternatively, one may experiment with extending the 2D CNN structure to 3D, by adding time as an extra dimension\cite{Ji2013,SottoVoce,predicting-3D}. Here, we follow the latter approach, and we investigate the applicability of a special 3D CNN model called the (2+1)D CNN~\cite{Tran} for ultrasound-based direct speech synthesis, and compare the results with those of a CNN+LSTM model. We find that our 3D CNN model achieves a lower error rate, while it is smaller, and its training is faster. We conclude that for ultrasound video-based SSI systems, 3D CNNs are definitely a feasible alternative to recurrent neural models.

The paper is structured as follows. Section \ref{sec_CNN} gives a technological overview of the CNNs we are going to apply. In Section \ref{sec_data} we describe the data acquisition and  processing steps for the ultrasound videos and the speech signal. Section \ref{sec_exp} presents our experimental set-up. We present the experimental results and discuss them in Section~\ref{sec_res}, and the paper is closed with the conclusions in Section~\ref{sec_concl}.

\section{Convolutional Neural Networks for Video Processing}
\label{sec_CNN}

\begin{figure}[t]
\centering
\includegraphics[width=0.9\textwidth]{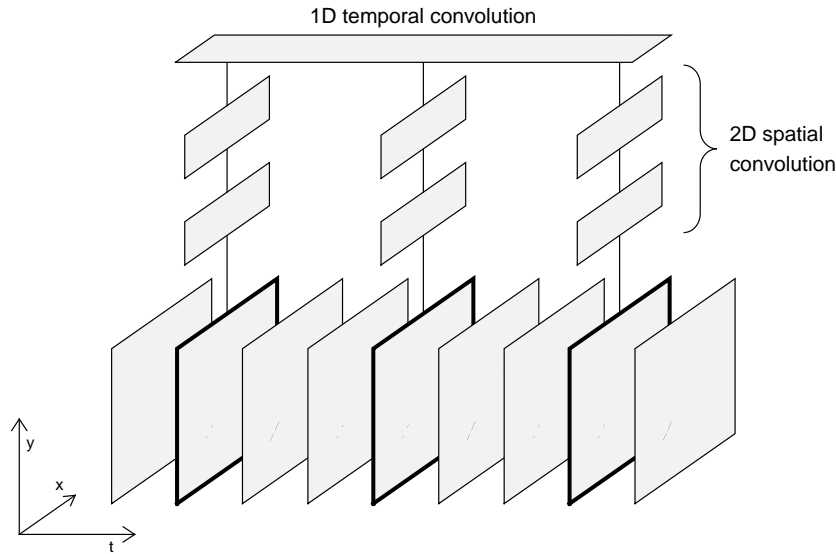}
\caption{Illustration of how the (2+1)D CNN operates. The video frames (at the bottom) are first processed by layers that perform 2D spatial convolution, then their outputs are combined by 1D temporal convolution. The model is allowed to skip video frames by changing the stride parameter of the temporal convolution.} \label{fig_cnn}
\end{figure}

Ever since the invention of 'Alexnet', CNNs have remained the leading technology in the recognition of still images~\cite{Imagenet}. These standard CNNs apply the convolution along the two spatial axes, that is, in two dimensions (2D). However, there are several tasks where the input is a video, and handling the video as a sequence (instead of simply processing separate frames) is vital for obtaining good recognition results. The best example is human gait recognition, but we can talk about action recognition in general~\cite{Ji2013,Zhao_pooling,Zhao_NIPS2018}.  In these cases, the sequence of video frames forms a three-dimensional data array, with the temporal axis being the third dimension in addition to the two spatial dimensions (cf. Fig.~\ref{fig_cnn}).

For the processing of sequences, recurrent neural structures such as the LSTM are the most powerful tool~\cite{hochreiter1997lstm}. However, the training of these networks is known to be slow and problematic, which led to the invention of simplified models, such as the gated recurrent unit (GRU)~\cite{cho2014learning} or the quasi-recurrent neural network~\cite{Bradbury}. Alternatively, several convolutional network structures have been proposed that handle time sequences without recurrent connections. In speech recognition, time-delay neural networks (TDNNs) have proved very successful~\cite{Peddinti2015,Toth-conv2}, but we can also mention the feedforward sequential memory network~\cite{FSMN}. 
As regards video processing, several modified CNN structures have been proposed to handle the temporal sequence of video frames~\cite{Ji2013,Zhao_pooling,Zhao_NIPS2018}. Unfortunately, the standard 2D convolution may be extended to 3D in many possible ways, giving a lot of choices for optimization. Tran et al. performed an experimental comparison of several 3D variants, and they got the best results when they decomposed the spatial and temporal convolution steps~\cite{Tran}. The model they called '(2+1)D convolution' first performs a 2D convolution along the spatial axes, and then a 1D convolution along the time axis (see Fig.~\ref{fig_cnn}). By changing the stride parameter of the 1D convolution, the model can skip
several video frames, thus covering a wider time context without increasing the number of processed frames. Interestingly, a very similar network structure proved very efficient in speech recognition as well~\cite{Toth-conv2}. Stacking several such processing blocks on top of each other is also possible, resulting in a very deep network~\cite{Tran}. Here, we are going to experiment with a similar (2+1)D network structure for ultrasound-based SSI systems.

\section{Data Acquisition and Signal Preprocessing}
\label{sec_data}

The ultrasound recordings were collected from a Hungarian female subject (42 years old, with normal speaking abilities) while she was reading sentences aloud. Her tongue movement was recorded in a midsagittal orientation -- placing the ultrasonic imaging probe under the jaw -- using a "Micro" ultrasound system by Articulate Instruments Ltd. The transducer was fixed using a stabilization headset. The 2-4 Mhz / 64 element 20 mm radius convex ultrasound transducer produced 82 images per second. The speech signal was recorded in parallel with an Audio-Technica ATR 3350 omnidirectional condenser microphone placed at a distance of 20 cm from the lips. The ultrasound and the audio signals were synchronized using the software tool provided with the equipment. Altogether 438 sentences (approximately half an hour) were recorded from the subject, which was divided into train, development and test sets in a 310-41-87 ratio. We should add that the same dataset was used in several earlier studies \cite{Csapo2017c,Grosz2018}. 

\begin{figure}[t]
\centering
\includegraphics[width=0.94\textwidth]{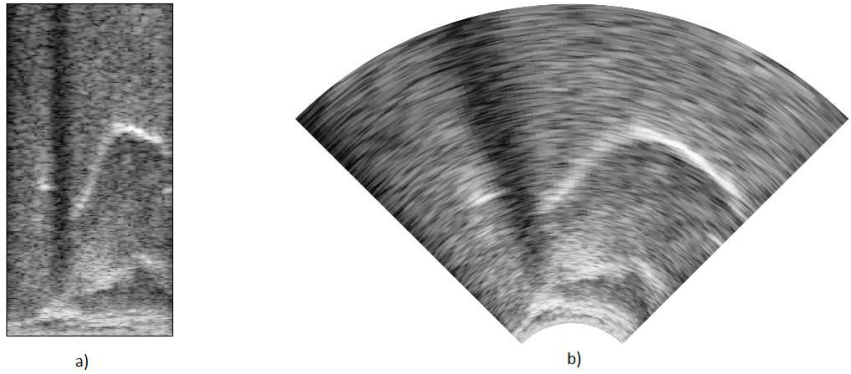}
\caption{Example of displaying the ultrasound recordings as a) a rectangular image of raw data samples b) an anatomically correct image, obtained by interpolation.} \label{fig1_us}
\end{figure}

The ultrasound probe records 946 samples along each of its 64 scan lines. The recorded data can be converted to conventional ultrasound images using the software tools provided. However, due to its irregular shape, this image is harder to process by computers, while it contains no extra information compared to the original scan data. Hence, we worked with the original 964x64 data items, which were downsampled to 128x64 pixels. Fig.~\ref{fig1_us} shows an example of the data samples arranged as a rectangular image, and the standard ultrasound-style display generated from it. The intensity range of the data was min-max normalized to the [-1, 1] interval before feeding it to the network. 

The speech signal was recorded with a sampling rate of 11025 Hz, and then processed by a vocoder from the SPTK toolkit (http://sp-tk.sourceforge.net). The vocoder represented the speech signals by 12 Mel-Generalized Cepstral Coefficients (MGCC) converted to a Line Spectral Pair representation (LSP), with the signal's gain being the 13th parameter. These 13 coefficients served as the training targets in the DNN modeling experiments, as the speech signal can be reasonably well reconstructed from these parameters. Although perfect reconstruction would require the estimation of the pitch (F0 parameter) as well, in this study we ignored this component during the experiments. To facilitate training, each of the 13 targets were standardized to zero mean and unit variance.

\section{Experimental Set-Up}
\label{sec_exp}

We implemented our deep neural networks in Keras, using a Tensorflow backend~\cite{chollet2015keras}. We created three different models: a simple fully connected network (FCN), a convolutional network that processes one frame of video (2D CNN), and a convolutional network that can process several subsequent video frames as input (3D CNN). To keep them comparable with respect to parameter count, all three models had approximately 3.3 million tunable parameters. Training was performed using the stochatic gradient descent method (SGD) with a batch size of 100.  The training objective function was the mean squared error (MSE).

\textbf{Fully Connected Network (FCN):} The simplest possible DNN type is a network with fully connected layers. To be comparable with an earlier study~\cite{Csapo2017c}, our FCN model consisted of 5 fully connected hidden layers, with an output layer of 13 neurons for the 13 training targets. The input of the network consisted of one video frame (128x64=8192 pixels). Each hidden layers had 350 neurons, so the model was about 4 times smaller compared to the FCN described in~\cite{Csapo2017c}. Apart from the linear output layer, all layers applied the swish activation function~\cite{swish}, and were followed by a dropout layer with the dropout rate set to 0.2.

\begin{table}[t]
\caption{The layers of the 2D and 3D CNNs in the Keras implementation, along with their most important parameters. The differences are highlighted in bold.}\label{tab1}
\centering
\renewcommand{\arraystretch}{1.1} 
\begin{tabular}{|l|l|}
\hline
\hfil2D CNN &  \hfil3D CNN \\
\hline
Conv2D(30, (13,13), strides=(2,2)) & Conv\textbf{3D}(30, (\textbf{5},13,13), strides=(\textbf{s}, 2,2))\\
Dropout(0.2) & Dropout(0.2)\\
Conv2D(60, (13,13), strides=(2,2)) & Conv\textbf{3D}(60, (\textbf{1},13,13), strides=(\textbf{1},2,2))\\
Dropout(0.2) & Dropout(0.2)\\
MaxPooling2D(pool\_size=(2,2)) & MaxPooling\textbf{3D}(pool\_size=(\textbf{1},2,2))\\
Conv2D(90, (13,13), strides=(2,1)) & Conv\textbf{3D}(90, (\textbf{1},13,13), strides=(\textbf{1},2,1))\\
Dropout(0.2) & Dropout(0.2)\\
Conv2D(120, (13,13), strides=(2,2))~~~& Conv\textbf{3D}(\textbf{85}, (\textbf{1},13,13), strides=(\textbf{1},2,2))~~~\\
Dropout(0.2) & Dropout(0.2)\\
MaxPooling2D(pool\_size=(2,2)) & MaxPooling\textbf{3D}(pool\_size=(\textbf{1},2,2))\\
Flatten() & Flatten()\\
Dense(500) & Dense(500)\\
Dropout(0.2) & Dropout(0.2)\\
Dense(13, activation='linear') & Dense(13, activation='linear')\\
\hline
\end{tabular}
\label{table_CNN}
\end{table}

\textbf{Convolutional Network (2D CNN):} Similar to the FCN, the input to this network consisted of only one frame of data. The network performed spatial convolution on the input image via its four convolutional layers below the uppermost fully connected layer. The actual network configuration is shown in Table~\ref{table_CNN}. The optimal network meta-parameters were found experimentally, and all hidden layers applied the swish activation function~\cite{swish}.

\textbf{3D Convolutional Network (3D CNN):} To enable the processing of video frames sequences, we changed the 2D convolution to 3D convolution in our CNN. This network processed 5 frames of video that were $s$ frames apart, where $s$ is the stride parameter of the convolution along the time axis. Following the concept of (2+1)D convolution described in Section~\ref{sec_CNN}, the five frames were first processed only spatially, and then got combined along the time axis just below the uppermost dense layer. Table~\ref{table_CNN} shows the actual network configuration. The modifications compared to the 2D CNN are shown in bold. We note that the number of filters in the uppermost convolutional layer was decreased in order to keep the number of parameters in the same range as that for the 2D CNN.

There are several options for evaluating the performance of our networks. In the simplest case, we can compare their performance by simple objective metrics, such as the value of the target function optimized during training (the MSE function in our case). Unfortunately, these metrics do not perfectly correlate with the users' subjective sense of quality of the synthesized speech. Hence, many authors apply subjective listening tests such as the MUSHRA method~\cite{Juanpere2019}. This kind of evaluation is tedious, as it requires averaging the scores of a lot of human subjects. As an interesting shortcut, Kimura et al. applied a set of commercial speech recognizers to substitute the human listeners in the listening tests \cite{SottoVoce}. In this paper, we will simply apply objective measures, namely the mean squared error (MSE) and the (mean) $R^2$ score, which are simple and popular methods for evaluating the performance of neural networks on regression tasks.

\section{Results and Discussion}
\label{sec_res}

As for the 3D CNN, we found that the value of the stride parameter $s$ has a significant impact on the error rate attained. The size of the input time context covered by the network can be calculated as $w=4\cdot s+1$. For example, for $s=6$ the distance between the first and the last time frames was $w=25$, meaning that the network input should consist of a sequence of 25 video frames. According to the 82 fps sampling rate, this corresponds to video chunks of about 300 ms. 

\begin{figure}[t]
\includegraphics[width=\textwidth]{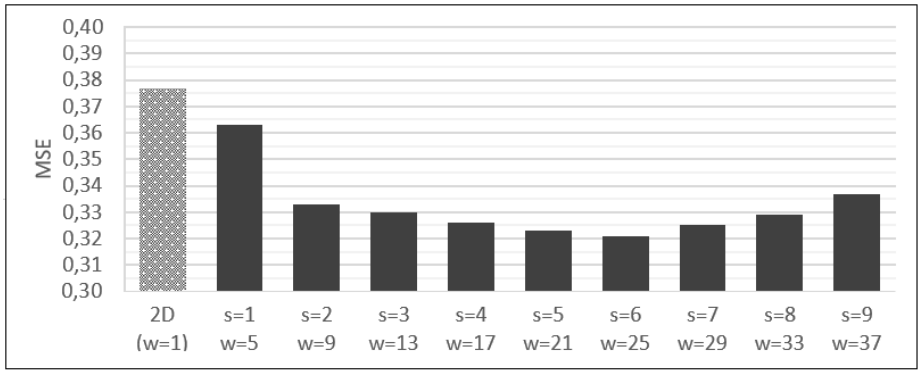}
\caption{MSE rates of the 3D CNN on the development set for various $s$ stride values. For comparison, the MSE attained by the 2D CNN is also shown (leftmost column).} \label{fig_3ds}
\end{figure}

Fig.~\ref{fig_3ds} shows the MSE obtained on the development set with the 3D CNN network for various $s$ values. As a base of comparison, the MSE attained by the 2D CNN that processes a single frame is also shown. It can be clearly seen that extending the actual frame with its context can significantly reduce the error rate. Including 2-2 immediate neighbors is already effective (s=1), but the largest gain was achieved when setting $s$ to a value between 3 and 8.

Table~\ref{table_res} summarizes the best results for the three network configurations, both for the development and test sets. Along with the MSE values, now the correlation-based $R^2$ scores are also shown. The 2D CNN network was superior to the fully connected network, but the 3D CNN clearly outperformed it, reducing the MSE on the test set by about 21\% in a relative sense. The $R^2$ score increased by 14\% absolute.

The two bottom rows of Table~\ref{table_res} compare our results with two earlier studies. The authors of~\cite{Csapo2017c} applied a fully connected network on the same data set. They obtained slightly better results than those given by our FCN, presumably due to the fact that their network had about 4 times as many parameters. More interestingly, they attempted to include more neighboring frames in the processing, simply by concatenating the corresponding image data. Feature selection was applied to alleviate the problems caused by the large size of the images ($\sim8000$ pixel per image), These simple methods failed to significantly reduce the error rate. Our current experiments show that the frames should be placed farther apart ($3\leq s\leq 8$) for optimal performance. Moreover, instead of reducing the input size by feature selection, it seems to be more efficient to send the frames through several neural layers, with a relatively narrow 'bottleneck' layer on top.

Moliner and Csapó combined the 2D CNN with an LSTM, this way processing video chunks of 32 frames~\cite{Juanpere2019}. As we explained in Section~\ref{sec_CNN}, this is the most viable and competitive alternative to our approach. Unfortunately, their study reports no MSE scores, but they provided us with their code. We retrained their (uni-directional) LSTM with some minimal and unavoidable modifications, e.g. adjusting the input and output layer sizes (they used different training targets and a slightly different input resampling). Their model had five times a many tunable parameters than our models, and its training also took much longer. While it clearly outperformed both the FCN and the 2D CNN models, it could not compete with our 3D CNN, in spite of its larger complexity.

\begin{table}[t]
\caption{The results obtained with the various network configurations. For comparison, two results from the literature are also shown in the bottom rows.}\label{tab2}
\centering
\renewcommand{\arraystretch}{1.2} 
\begin{tabular}{|c||c|c||c|c|}
\hline
Network &  \multicolumn{2}{|c||}{Dev} & \multicolumn{2}{c|}{Test} \\
\cline{2-5}
Type &  ~~~MSE~~~ & ~~~Mean $R^2$~~~ &  ~~~MSE~~~ & ~~~Mean $R^2$~~~\\
\hline
~~~FCN & 0.408 & 0.599 & 0.400& 0.598 \\
\hline
~~~2D CNN & 0.377 & 0.630 & 0.366 & 0.633 \\
\hline
~~~3D CNN (s=6)~~~ & 0.321 & 0.684 & 0.315 & 0.683 \\
\hline
\hline
~~~FCN~\cite{Csapo2017c} & 0.384 & 0.619 & n/a & n/a \\
\hline
~~~CNN + LSTM~\cite{Juanpere2019}~~~ & 0.345 & 0.653 & 0.336 & 0.661 \\
\hline
\end{tabular}
\label{table_res}
\end{table}

\section{Conclusions}
\label{sec_concl}

Here, we implemented a 3D CNN for ultrasound-based articulation-to-acoustic conversion, where the CNN applied separate spatial and temporal components, motivated by the (2+1)D CNN of Tran et al.~\cite{Tran}. The model was compared with a CNN+LSTM architecture that was recently proposed for the same task. We found that the 3D CNN performed slightly better, while it was smaller and faster to train. Though asserting the superiority of the 3D CNN would require more thorough  comparisons, we can safely conclude that 3D CNNs are viable competitive alternatives to CNN+LSTMs for the task of building SSI systems based on ultrasound videos of the tongue movement. In the future, we plan to investigate more sophisticated network types such as the ConvLSTM network that directly integrates the advantages of the convolutional and LSTM units~\cite{Predicting-ConvLSTM}.

\section{Acknowledgements}

This study was supported by the National Research, Development and Innovation Office of Hungary through project FK 124584 and by the AI National Excellence Program (grant 2018-1.2.1-NKP-2018-00008) and by grant TUDFO/47138-1/2019-ITM of the Ministry of Innovation and Technology. László Tóth was supported by the UNKP 19-4 National Excellence Programme of the Ministry of Innovation and Technology, and by the J\'anos Bolyai Research Scholarship of the Hungarian Academy of Science. The GPU card used for the computations was donated by the NVIDIA Corporation. We thank the MTA-ELTE Lendület Lingual Articulation Research Group for providing the ultrasound recordings.

%
%
%
 \bibliographystyle{splncs04}
 \bibliography{ICAISC2020}
%

\end{document}